\documentclass[conference]{IEEEtran}
\IEEEoverridecommandlockouts
\usepackage{cite}
\usepackage{amsmath,amssymb,amsfonts}
\usepackage{algorithmic}
\usepackage{graphicx}
\usepackage{textcomp}
\usepackage{xcolor}
\usepackage[nolist]{acronym}
\def\BibTeX{{\rm B\kern-.05em{\sc i\kern-.025em b}\kern-.08em
    T\kern-.1667em\lower.7ex\hbox{E}\kern-.125emX}}
\begin{document}

\title{Experimental Demonstration of Network Convergence with Coherent and Analog Radio-over-Fibre signals For Densified 5.5G/6G Small Cell Networks\\
}

\author{  
Frank Slyne\IEEEauthorrefmark{1}, Colm Browning\IEEEauthorrefmark{2}, Amol Delmade\IEEEauthorrefmark{2}, Liam P. Barry \IEEEauthorrefmark{2} and Marco Ruffini\IEEEauthorrefmark{1}\\ 
\IEEEauthorblockA{\IEEEauthorrefmark{1} CONNECT centre, School of Computer Science and Statistics, Trinity College Dublin (TCD), Ireland. \\ \IEEEauthorrefmark{2} School of Engineering, Dublin City University (DCU), Ireland \\ (slynef@tcd.ie, colm.browning@dcu.ie, amol.delmade2@mail.dcu.ie, liam.barry@dcu.ie, marco.ruffini@tcd.ie)}
}

\maketitle

\begin{abstract}
In this work we analyse and demonstrate the coexistence of digital coherent and analogue radio over fibre signals over an access-metro transmission network and field fibre. We analyse how the spectral proximity of the two signals and the non-ideal filter alignment of typical telecomms-grade ROADMs affect the signal performance. Our results show that coexistence is indeed possible, although performance deteriorates with the increase in number of ROADMs in the network topology. Thus, while today's access-metro networks will be able to support future 5.5 and 6G cell densification operating at mmWave and THz frequency, using spectral efficient analogue radio over fibre transmission, there will be trade-offs to be considered. In our experiment setup, we show that the limit for ARoF accessible performance is reached after transmission over 3 ROADMs and a total of 49 km of fibre.  

\end{abstract}







\section{Introduction}

The drive towards increasing densification of mobile network technology, has led to much research on cost-effective fronthaul and backhaul technology \cite{Raddo:19, Browning}. In addition to the large increase in capacity expected in the coming years, the adoption of millimeter-wave (mmWave) and Terahertz (THz) transmission technology will require site densification due to the limited reach at these high carrier frequencies \cite{Huq:19}. As 5.5 and 6G will develop higher and higher capacity, the required transport data rate on the optical fibre will continue to increase \cite{PfeifferJLT}. This is exacerbated by the use of RAN disaggregation \cite{AlvarezOFC}, functional split and centralisation of processing \cite{RuffiniJLT}, which further increases capacity and latency constraints on the fibre transmission \cite{Saliou:21}. While transport of fronthaul signals is still feasible for todays' commercial base station capacity, the move towards much higher wireless bandwidth at the base stations (e.g., of 1 GHz and above) for mmWave/THz communications, will require high-cost transmission technology to support data rates of hundreds Gb/s. In these scenarios, Analogue Radio over Fibre (ARoF) transmission can satisfy the need to transport high-bandwidth radio signals over fibre in a way that is both cost effective and spectrally efficient \cite{ARoFBook}.

With A-RoF fronthaul, the wireless signal is transmitted from the Distributed Unit (DU), which may be co-located with optical networking hardware at a \ac{CO}, through a fibre link or network to a remotely deployed radio Unit (RU). The multi-carrier radio signal is transmitted on the RF carrier or on an Intermediate Frequency (IF) (known as IF A-RoF). Transmission in this manner allows the removal of costly \acp{DAC} and \acp{ADC} from the RU, reducing hardware complexity. This is highly advantageous in terms of operator \ac{CAPEX}, and thereby helps to facilitate cell site densification into the 6G era. 

Compared to \ac{D-RoF} fronthaul, \ac{A-RoF}  offers higher spectral efficiency, as there is no requirement for quantization of the \ac{RF} signals. The inherent high \ac{PAPR} of multi-carrier modulation methods such as \ac{OFDM}, widely employed in wireless technologies, leads to a limitation in terms of optical power budget, ultimately dictating the transmission distance and/or number of serviced cells. \ac{A-RoF} signals are known to be significantly degraded in the presence of non-linearities, which are typical of optical fibre transmission \cite{Novak:16}. However, as wide-scale use of shorter range mmWave/THz communications is expected, in order to provide ever higher wireless capacity, the motivations for ARoF deployment are stronger than ever. The use of IF-ARoF requires the RU to upconvert the received IF signal to the required RF for wireless transmission, but provides spectral flexibility in the optical domain - a property which can be advantageous where ARoF is provisioned over a shared optical transport network, e.g. with Wavelength Division Multiplexing (WDM) technology. This concept, known as network convergence, enables RoF provisioning to not only exploit pre-existing fibre infrastructure (such as metro and access topologies) for radio access transport, but also to avail of advances in optical switching/routing technologies deployed in state-of-the-art optical networks. Optimising the use of existing assets through infrastructure sharing, providing higher bandwidth, new degrees of optical flexibility to the radio domain and making use of highly spectrally efficient ARoF technologies can be leveraged to form a truly sustainable platform for radio access bandwidth and connectivity growth. But, to date, studies of converged radio fronthaul and optical metro/access transport are largely confined to relatively simplistic laboratory-based transmission experiments, often focusing on niche applications.

Successful convergence of ARoF signals at 5 GHz with a single carrier 10 Gb/s 4-level Pulse Amplitude Modulation (PAM) optical access signal in a passive optical network system scenario was demonstrated in \cite{Browning:17}. Maintaining a focus on access domain convergence, \cite{Li:21} demonstrates 56 Gb/s PAM-4 optical broadband signal transmission in tandem with 10 $\times$ 400 MHz multi-carrier signals at 28 GHz for mmWave ARoF transport. A network convergence experiment was carried out in \cite{Kanta:22}, where the authors demonstrated a dynamically reconfigurable switching node with coherent and ARoF signals. The node, based on Wavelength Selective Switch (WSS) technology, routes Digital Radio-over-Fiber and  Analog–Intermediate-Frequency over Fiber (A-IFoF) to support densified 5G small-cell. However the experiment was carried out using laboratory testbench equipment. Other work on fixed-mobile convergence in \cite{Munoz} focused in stead on archictecture and SDN-driven orchestration.

Our experiment demonstrates the coexistence of digital coherent and ARoF signal, using equipment that is typical of today's operator networks. A-IFoF coexists in close proximity to coherent signals in common channels providing efficient and spectrum utilisation that complies with Fixed Grid scheme.  
The experiment uniquely combines optical networking equipment located at the Open Ireland testbed in Trinity College Dublin (TCD), and hybrid RF/Optical research facilities located in a different lab at Dublin City University (DCU). 



\section{Experiment}
The use case we target is that of high densification of small cells for 5.5 and 6G technology, where some of the terminations are connected through ARoF, while others make use of digital transmission using standard coherent transmission technology (i.e. for signals of 100Gb/s and above). The key point is however that the ARoF and other coherent signals make use of existing metro network infrastructure to be carried towards their destination (e.g., a telco cloud). This is important in order to minimise the capital expenditure required for the back haul network. In addition, different signals might be multiplexed at different locations, as the metro network will carry several signals of varying bandwidth across the given area.
The reason we focus on ARoF transmission is that these signals have much higher spectral efficiency compared to digital RoF transmission, thus they can occupy a minimal part of the fibre capacity, which is especially important as we move to large wireless channel bandwidths from hundreds of MHz to a few GHz (i.e., for Thz communications) \cite{PfeifferJLT}.
The target of our analysis in to understand how the ARoF signal deteriorates when it is transmitted over a metro networks, across multiple ROADMs (with typical impairments found in telecomms-grade equipment), and understand how the Error Vector Magnitude (EVM) for the analogue signal increases as the signal moves through the network.

In our experiment, a wireless baseband signal of 488.28 Mhz bandwidth is first modulated at Intermediate Frequency (IF) of 1.5 GHz. This signal is then used to modulate an optical carrier, which is transmitted over fibre.
The signal is injected into a network of multiple ROADMs, where a 100Gb/s DP-QPSK coherent signal is added to the system and co-propagates with the ARoF signal.
The coherent signal operates at a rate of 31.5 Gbaud, for which a channel of 37.5 GHz is allocated in the flexgrid ROADMs. The ARoF Dual Sideband (DSB) signal we used only occupies 3.488 GHz (carrier plus 2 sidebands), and is allocated to a 6.25 GHz slot in the ROADM (this is the smallest granularity that can be allocated in typical telecomms-grade flexgrid ROADMs).

\subsection{Experiment setup}

\begin{figure*}
  \includegraphics[width=\textwidth,height=6cm]{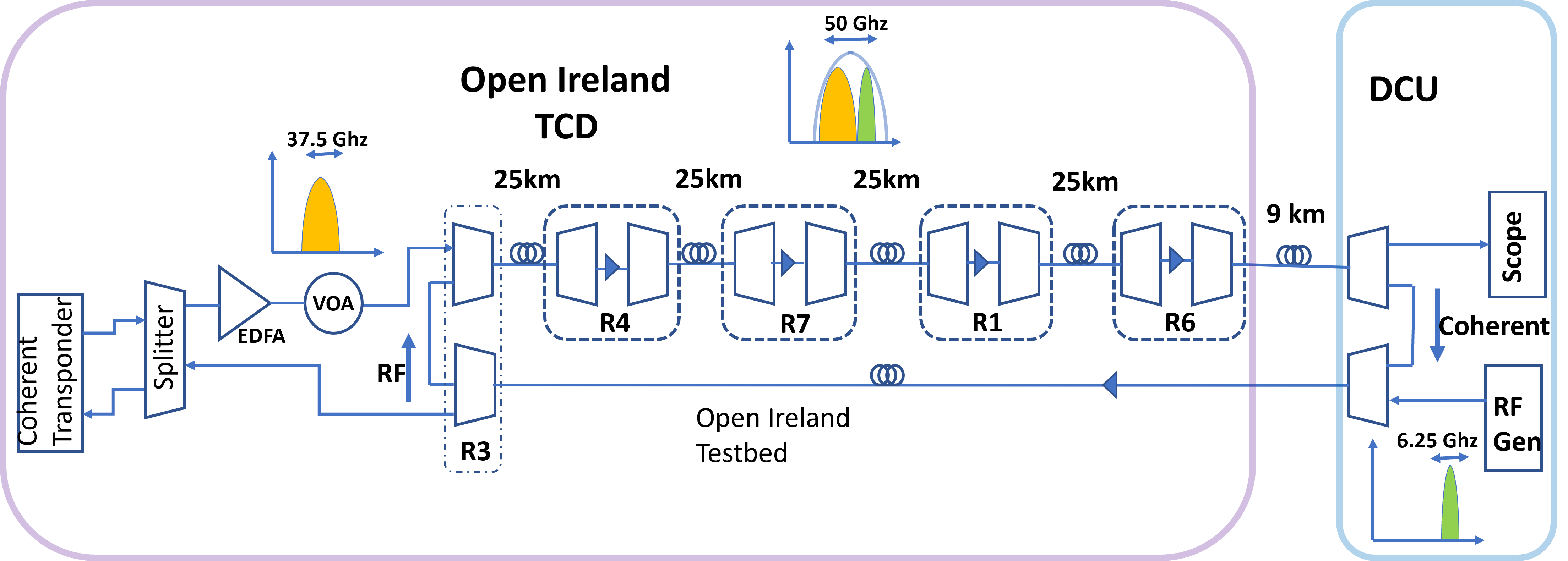}
  \caption{End to end architecture of the network convergence experiment}
  \label{fig:architecture}
\end{figure*}

\begin{figure*}
  \includegraphics[width=\textwidth]{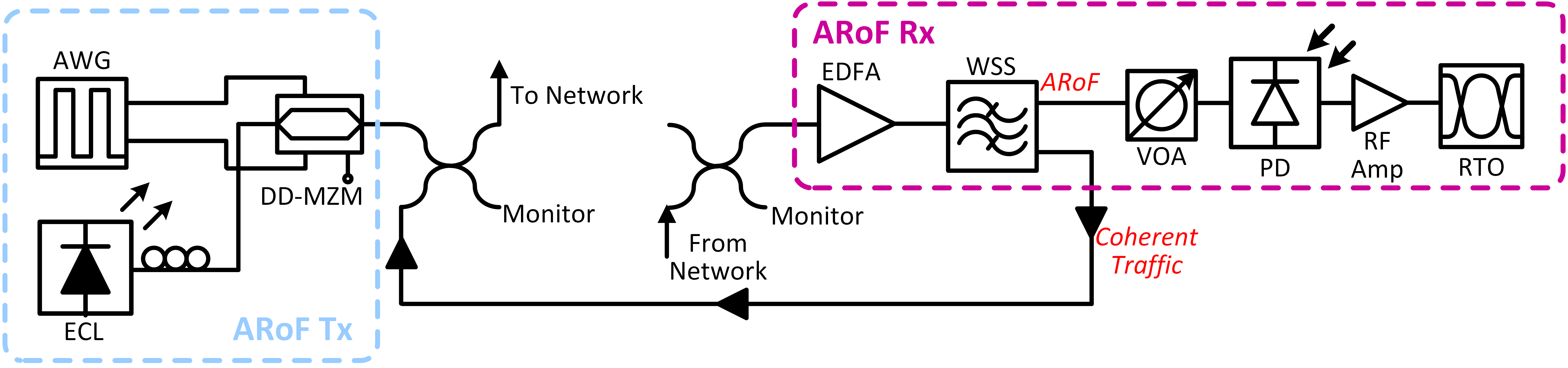}
  \caption{ARoF Tx/Rx experiment setup}
  \label{fig:ARoFarchitecture}
\end{figure*}

The experimental platforms used in this proposal are the OpenIreland testbed located in Trinity College Dublin and the Dublin City University (DCU) Radio and Optical Communications Laboratory. The two are linked through a dark fibre link (field fibre) across the city of Dublin of 9 km length.

The coherent transponders, ROADMs and fibre spools are located in the OpenIreland testbed. The transponder are ADVA Teraflex FSP3000C, that can operate at variable baud rate (31.5 - 71.96 Gbaud) and modulation schemes (DP-QPSK - DP-64QAM) to achieve bit rates from 100G through to 600Gbps. In this experiment, we use the most basic configuration DP-QPSK at 31.5 Gbaud and use a signal bandwidth of 37.5 Ghz. The line rate is 100 Gbps. 
The system architecture is shown in Fig. \ref{fig:architecture}. The coherent transponder is connected to an 8-way passive splitter to allow for signal monitoring (i.e., through an OSA) and diagnostics. A booster EDFA amplifies the signal before transmission. This is followed by VOA to allow dynamic selection of the output power.
The overall system is organised in a looped topology. The ARoF signal is generated at DCU and sent to TCD, to then return to DCU for decoding after passing through a number of ROADMS. The Coherent signal is generated at TCD, propagated across the ROADMs, then sent to DCU and form there back to TCD to be decoded.

More in detail, the digital coherent signal is injected to the MUX of ROADM3 (R3), which performs the function of discriminating and combining it with the ARoF signals (generated at DCU). An Add rule is applied the Mux WSS to add a 37.5 GHz wide band (194931.25 to 194968.75 GHz) for the coherent signal. An Add rule is also applied to the Mux WSS for a 6.25 GHz wide band (194968.75 to 194975.00 GHz) for the RF signal. Equivalent rules are added to the Demux side: the coherent signal is dropped towards the coherent receiver, while the ARoF signal is sent back to DCU after passing through the chain of ROADMs.

The chain of ROADMs is composed by a series of two-degree ROADMs connected through 25 km of SMF (notice that in this experiment we only consider a single direction of communication). The number of traversed ROADMs varies during the experiment to analyse how this affects the signal quality. After the last ROADM, the signal goes through 9 km of dedicated dark fibre, leased from a telecommunications operator, to reach the DCU Radio and Optical Communications Laboratory. 


Fig. \ref{fig:ARoFarchitecture} shows the laboratory experimental setup for the ARoF generation, transmission and reception at DCU. An Intermediate Frequency (IF) Orthogonal Frequency Division Multiplexing (OFDM) with 64-QAM is generated offline using matlab and loaded into an Arbitrary Waveform Generator (AWG) operating at 60 GSa/s. The 2.93 Gb/s ARoF signal, whose numerology is shown in Table \ref{tab:Num}, is output from the AWG and used to drive a Dual-Drive Mach Zehnder Modulator (DD-MZM) that modulates the light from an External Cavity Laser (ECL) operating at 194.97 THz (1537.6 nm). The output double side band optical signal is combined with the coherent DP-QPSK signal originating from TCD via a 50/50 optical coupler and the converged ARoF/Coherent services are transmitted at a total launch power of +7 dBm over one fibre of the dark fibre pair, towards TCD (denoted `To Network' in Fig. \ref{fig:ARoFarchitecture}).

\begin{table}[h]
\caption{ARoF signal numerologies.}
\label{tab:Num}
    \centering
\begin{tabular}{l|c|c}
\hline\noalign{\smallskip}
\textbf{Property} & \textbf{Value} & \textbf{Unit}\\
\noalign{\smallskip}\hline\noalign{\smallskip}
\# Subcarriers  & 250  &  n/a\\
(I)DFT size  & 1024 &  n/a\\ 
QAM order & 64 &  n/a \\
IF & 1.5 & GHz \\
Symbol Rate  & 1.95 &  MHz\\ 
Bandwidth & 488.28 &  MHz\\ 
Cyclic Prefix  & 6.25 & \% \\
Raw Data Rate  & 2.93 &  Gb/s\\
\hline \noalign{\smallskip}
\end{tabular}
\end{table}

\begin{figure}[h]
\centering
\includegraphics[width=1\linewidth]{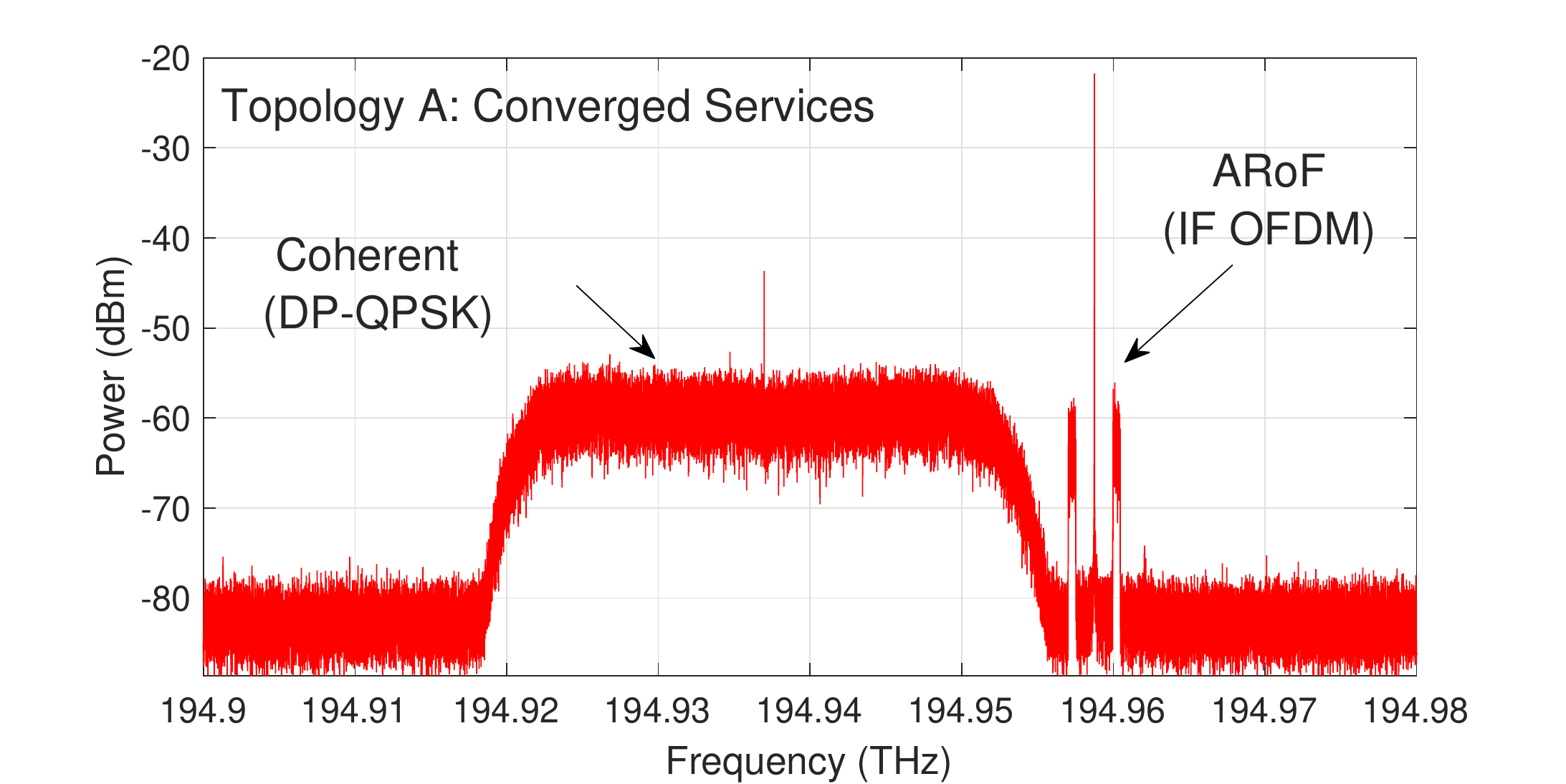}
\caption{\label{fig:ConvergedSpectrum} Optical spectrum of the digital coherent signal and analogue ROF signal combined, measure in the DCU lab.}
\end{figure}

The ARoF receiver at DCU receives the multiplexed ARoF/coherent optical signals from TCD via the second fibre of the dark fibre pair (denoted `From Network' in Fig. \ref{fig:ARoFarchitecture}). An example of the optical signal spectrum captured at this point is shown in Fig. \ref{fig:ConvergedSpectrum}. This received signal is first amplified using an in-line EDFA and then sent to a WSS, whose programmable filter profile is set to discriminate between the adjacent services, directing the ARoF signal for local detection via one output port and recirculating the coherent signal back to TCD via a second port as shown in Fig. \ref{fig:ARoFarchitecture}. A Variable Optical Attenuator (VOA) is used to set the ARoF optical power incident on a 20 GHz Photo-Detector (PD). The output electrical signal is amplified and sampled using a Real Time Oscilloscope operating at 50 GSa/s. The digitized ARoF signal is then processed offline including IF down-conversion, demodulation and Error Vector Magnitude (EVM) measurement. 

\begin{figure*}[h]
\centering
\includegraphics[width=0.8\linewidth]{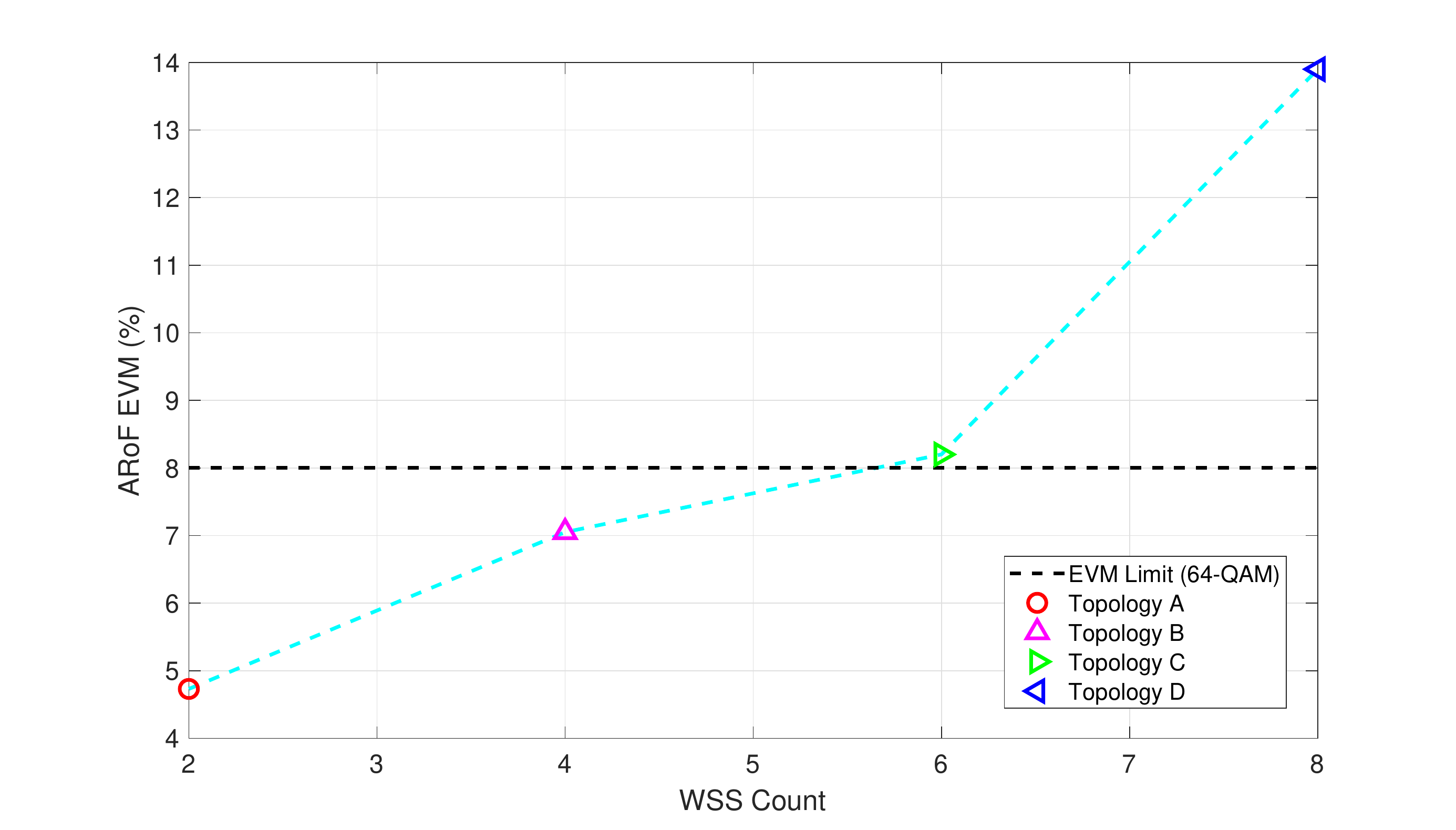}
\put(-355,70){\includegraphics[width=3cm]{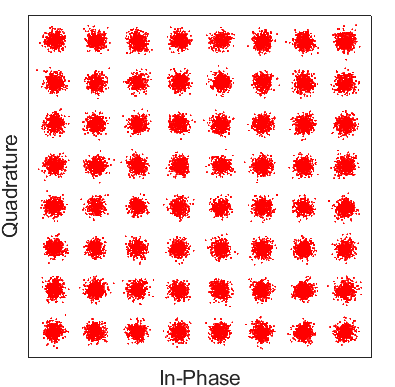}}
\put(-220,110){\includegraphics[width=3cm]{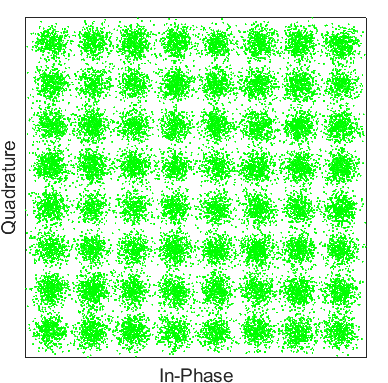}}
\caption{\label{fig:EVM} Error Vector Magnitude of the ARoF signal, for different propagation distance and WSS counts (calculated across the different topologies A, B, C and D). EVM 8\% limit for LTE highlighted.}
\end{figure*}

\begin{figure}[h]
\centering
\includegraphics[width=1\linewidth]{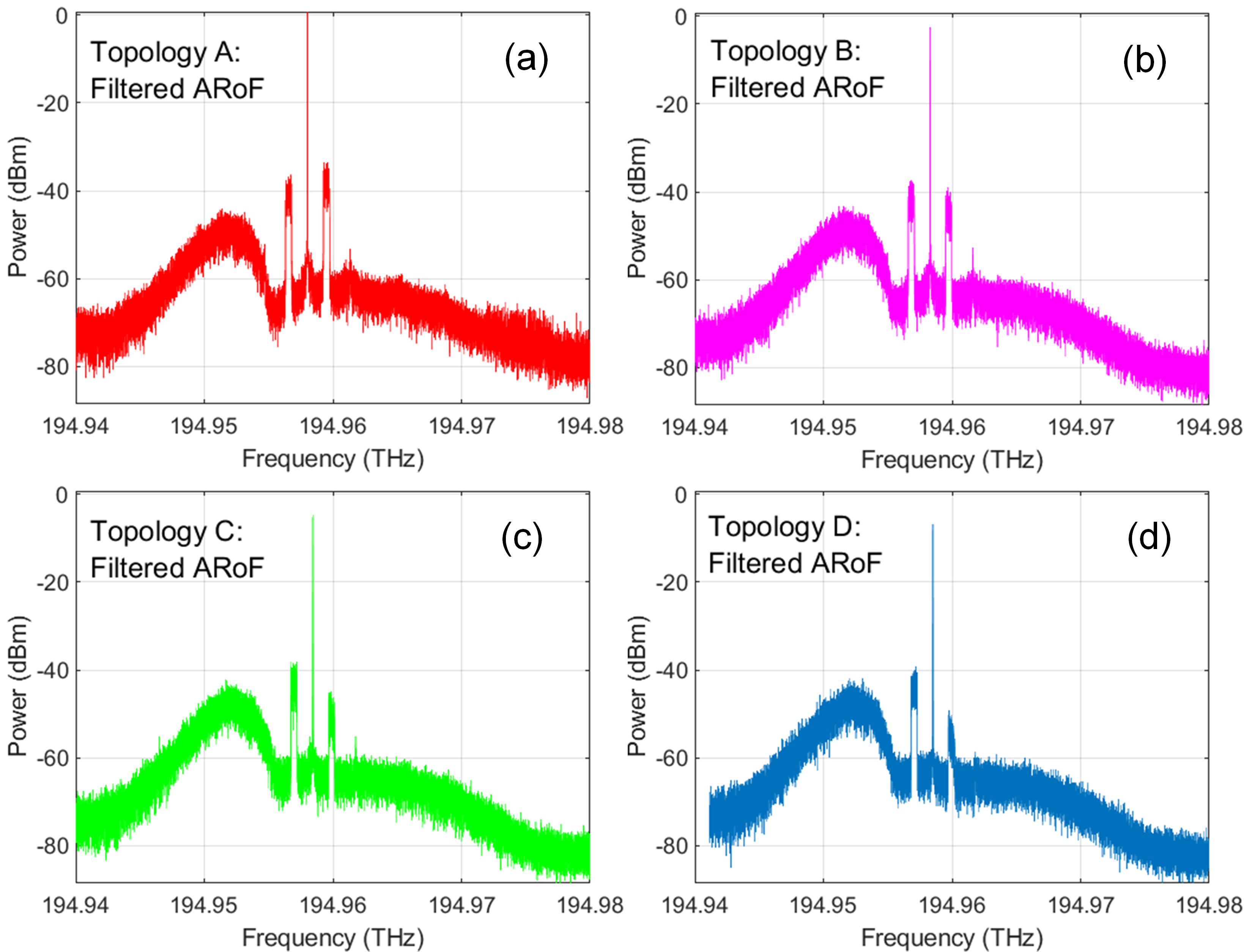}
\caption{\label{fig:Spectra} Optical spectra showing the DSB ARoF signal at the receiver for the different topologies A, B, C and D.}
\end{figure}

\subsection{Procedure}
To conduct the experiment, we first configure the physical lightpath topology in Open Ireland and DCU. In Open Ireland, since all components are routed through a large port-count fibre switch, physical paths can be set up programmatically. This enables fast and easily reproducible experiment setup, in addition to monitoring of power levels at all points of the topology, to easily assess any change in power (all ports of the fibre switch have power monitoring). Next, we configure the optimal transmit power levels of the transponders, and the gain of each EDFA and attenuator in the topology. Within each ROADM, the Mux WSS is followed by a booster EDFA, with gain configurable between 5 and 25 dB. The Demux WSS is instead preceded by an EDFA preamp, with gain also configurable between 5 and 25 dB. Each WSS add/drop port has a VOA with attenuation configurable between 0 and 20 dB. Lastly, we iterate through 4 different topologies, referred to with A, B, C and D, with varying number of ROADMs, as shown in Table \ref{fig:topology}. 

\begin{table}[]
\caption{\label{fig:topology} Topologies used, with associated number of WSS and overal transmission distance.}
 \centering
\begin{tabular}{|l|l|l|}
\hline
\textbf{Topology} & \textbf{WSS Count} & \textbf{\begin{tabular}[c]{@{}l@{}}Fiber Length\\ km\end{tabular}} \\ \hline
A & 2 & 9 \\ \hline
B & 4 & 34 \\ \hline
C & 6 & 59 \\ \hline
D & 8 & 84 \\ \hline
\end{tabular}
\end{table}

Each ROADM (1x20 Lumentum ROADM Graybox) has 2 WSS's (one for MUX and one for DEMUX). For the first topology (A), we only use ROADM3 in TCD and the WSS in DCU that discriminates the Coherent and RF signals. For the second topology (B), a further ROADM is added in (thus, 2 more WSS), along with a length of 25 km fibre. For the third topology (C), we have 3 ROADMs (6 WSS), and finally, in the fourth topology (D), 4 ROADMs (8 WSS) are used. The shortest combined length (i.e., across the field fibre and lab reels) is 9 km, while the longest is 84 km.

\section{Results and Discussion}
The performance metrics we collect for the digital coherent transmission are the real-time values for optical received power, Q-Factor, signal to noise ratio, optical signal to noise ratio, pre-FEC bit error rate and the uncorrected blocks, all of which are reported directly from the transponder. Since the DP-QPSK signal is resilient to noise and has typically a reach above 1000km, it’s performance is not affected by the ARoF signal and are similar across all tested topologies, with Q-factor of 16.5, SNR around 18.5 dB and OSNR of around 28.5 dB, and negligible pre-FEC BER.



For the ARoF 64-QAM OFDM signal we instead measure the Error Vector Magnitude (EVM), shown in Fig. \ref{fig:EVM} for the different topologies. The x axis of the figure indicates the overall number of WSSs traversed. In the simplest transmission case (Topology A) the received ARoF signal exhibits excellent performance with an EVM of 4.7\%. The received ARoF optical spectrum measured for this topology, after signal discrimination, is shown in Fig. \ref{fig:Spectra}(a). The figure shows the DSB ARoF signal alongside a portion of the DP-QPSK coherent signal which is present due to spectral leakage owing to the proximity of the signals as they pass through the non-ideal filter of the WSS signal discriminator. Nevertheless, satisfactory performance is achieved with the received constellation shown as an inset in Fig. \ref{fig:EVM} where the received optical power was 2 dBm.

The figure also shows ARoF degradation when the WSS count increases, as the signal propagates through additional filters and longer distance. This degradation can be understood by observing the respective measured ARoF optical spectra for each topology in Figs. \ref{fig:Spectra}(a), (b), (c) and (d). The spectra reveal that transmission through additional ROADM filters results in increased attenuation of the ARoF signal's upper sideband. This effect occurs due to the non-perfect alignment of WSS filters in each ROADM and specifically limits ARoF performance. The net result is an attenuation of the ARoF signal, leading to lower received optical powers and a degradation in signal-to-noise ratio. Fig. \ref{fig:Spectra} shows EVMs of 7.05\%, 8.2\% and 13.9\% for topologies B, C and D, with these performances measured with received optical powers of  -1 dBm, -3 dBm and -6 dBm, respectively. The figure also shows the received 64-QAM constellation measured after transmission over topology C as an inset.

These results are highly valuable as they reveal a new practical limitation on the performance of ARoF transmission in the case of convergence with flexgrid coherent services and networking equipment. The degradation highlighted by the results could be alleviated through tailoring of the channel spacing dedicated for adjacent ARoF fronthaul and coherent metro traffic, or altering the format or bandwidth of the wireless signal, indicating an interesting trade-off between spectral efficiency and the required complexity of the fronthaul transmission path. The choice of IF for ARoF may also be varied here to account for sideband attenuation. In the scenarios evaluated here, the results shows that ARoF and coherent signal convergence is indeed a viable networking option for future heterogenous optical networks, with results for topology C (6 WSSs) exhibiting performance just at the current limit for 64-QAM OFDM 5G NR signals.

\section{Conclusions}
In this work we have analysed the coexistence of digital coherent and analogue radio over fibre signals in an access-metro network architecture. Our experimental setup included telecomms-grade equipment (ROADMs, digital transponders and amplifiers) and a 9 km field fibre (linking TCD and DCU Universities). Transmission windows of 37.5 GHz and 6.25 GHz, respectively, for the 31.5Gbaud digital coherent DP-QPSK and the 64-QAM OFDM ARoF signals, where created by the ROADMs. In order to assess the degradation of the signals across the metro networks, mostly generated by the non-ideal filter alignment of different ROADM WSS, we have created 4 different topologies with increasing number of ROADMs and transmission distance, to recreate a typical metro transmission network. Our results show that filter misalignment that is typical of telecomms-grade ROADMs does affect the ARoF signal quality, limiting the transmission to a topology with 3 ROADMs. This demonstrates that while transmission of ARoF signal is feasible in toady's access-metro networks, there are trade-offs that needs to be considered for topologies with higher number of ROADMs. In our future work we intend to further investigate these trade-offs, analysing the behaviour of ARoF signals with different bandwidth and modulation formats, the use of ROADM windows with different bandwidth and the effect of different channel loading configurations (i.e., loading the system with a higher number of digital coherent channels).


\section*{Acknowledgments}
This work has been funded by DTIF  EI grant DT 2019 0014B (Freespace), Science Foundation Ireland projects OpenIreland 18/RI/5721, 13/RC/2077\_p2 and 18/SIRG/5579. We acknowledge the technology support of ADVA.

\begin{acronym} 

\acro{A-RoF}{Analog Radio-over-Fiber}
\acro{ABNO}{Application-Based Network Operations}
\acro{A-CPI}{Application Controller Plane Interface}
\acro{ADC}{Analog-to-Digital Converter}
\acro{API}{Application Programming Interface}
\acro{AWG}{Arrayed Waveguide Grating}
\acro{B-RAS}{Broadband Remote Access Servers}
\acro{BB}{Base Band}
\acro{BBU}{Base Band Unit}
\acro{BER}{Bit Error Rate}
\acro{BGP}{Border Gateway Protocol}
\acro{BS}{Base Station}
\acro{BtB}{Back to Back}
\acro{BW}{Bandwidth}
\acro{CAP}{Carierless Amplitude Phase}
\acro{C-RAN}{Cloud Radio Access Networks}
\acro{C-RoFN}{Cloud-based Radio over optical Fiber Network}
\acro{CAPEX}{Capital Expenditure}
\acro{CDN}{Content Distribution Network}
\acro{CDR}{Clock Data Recovery}
\acro{CIR}{Committed Information Rate}
\acro{CO}{Central Office}
\acro{CoMP}{Coordinated Multipoint}
\acro{COP}{Control Orchestration Protocol}
\acro{CPE}{Customer Premises Equipment}
\acro{CPRI}{Common Public Radio Interface}
\acro{CS}{Central Station}
\acro{D-RoF}{Digital Radio-over-Fiber}
\acro{DAC}{Digital-to-Analog Converter}
\acro{DAS}{Distributed Antenna Systems}
\acro{DBA}{Dynamic Bandwidth Allocation}
\acro{DC}{Direct Current}
\acro{DFB}{Distributed Feedback Laser}
\acro{D-CPI}{D Controller Plane Interface}
\acro{DL}{Downlink}
\acro{DMT}{Discrete Multitone}
\acro{DSB}{Double Side Band}
\acro{DSLAM}{Digital Subscriber Line Access Multiplexer}
\acro{DSL}{Digital Subscriber Line}
\acro{DSP}{Digital Signal Processing}
\acro{DSS}{Distributed Synchronization Service}
\acro{DWDM}{Dense Wave Division Multiplexing}
\acro{E-CORD}{Enterprise CORD}
\acro{EDF}{Erbium-Doped Fiber}
\acro{eICIC}{enhanced Inter-Cell Interference Coordination}
\acro{EMBS}{Elastic Mobile Broadband Service}
\acro{EON}{Elastic Optical Network}
\acro{EPON}{Ethernet Passive Optical Network}
\acro{EPC}{Evolved Packet Core}
\acro{ETSI}{European Telecommunications Standards Institute}
\acro{EVM}{Error Vector Magnitude}
\acro{FBMC}{Filterbank Multicarrier}
\acro{FEC}{Forward Error Correction}
\acro{FDD}{Frequency Division Duplex}
\acro{FFR}{Fractional Frequency Reuse}
\acro{FFT}{Fast Fourier Transform}
\acro{FPGA}{Field-Programmable Gate Array}
\acro{FSO}{Free-Space-Optics}
\acro{FSR}{Free Spectral Range}
\acro{FTTB}{Fibre to the Building}
\acro{FTTcab}{Fibre to the CAB}
\acro{FTTH}{Fibre to the Home}
\acro{FTTx}{Fibre to the x}
\acro{G.Fast}{Fast Access to Subscriber Terminals}
\acro{GFDM}{Generalized Frequency Division Multiplexing}
\acro{GMPLS}{Generalized Multi-Protocol Label Switching}
\acro{GPON}{Gigabit Passive Optical Network}
\acro{GPP}{General Purpose Processor}
\acro{GPRS}{General Packet Radio Service}
\acro{GTP}{GPRS Tunneling Protocol}
\acro{HA}{Hardware Accelerator}
\acro{HARQ}{Hybrid-Automatic Repeat Request}
\acro{I2RS}{Interface 2 Routing System}
\acro{IA}{Interference Alignment}
\acro{ICT}{Information and Communication Technology}
\acro{I-CPI}{I Controller Plane Interface}
\acro{IETF}{Internet Engineering Task Force}
\acro{IF}{Intermediate Frequency}
\acro{IoT}{Internet of Things}
\acro{IP}{Internet Protocol}
\acro{IQ}{In-phase/Quadrature}
\acro{IRC}{Interference Rejection Combining}
\acro{KVM}{Kernel-based Virtual Machine}
\acro{LAN}{Local Area Network}
\acro{LO}{Local Oscillator}
\acro{LOS}{Line Of Sight}
\acro{LR-PON}{Long Reach PON}
\acro{LSP-DB}{Label Switched Path Database}
\acro{LTE}{Long Term Evolution}
\acro{LTE-A}{Long Term Evolution Advanced}
\acro{MAC}{Medium Access Control}
\acro{M-CORD}{Mobile CORD}
\acro{MEC}{Multi-Access Edge Computing}
\acro{MME}{Mobility Management Entity}
\acro{mmWave}{millimeter Wave}
\acro{MNO}{Mobile Network Operator}
\acro{MIMO}{Multiple Input Multiple Output}
\acro{MU-MIMO}{Multi-user MIMO}
\acro{MPLS}{Multiprotocol Label Switching}
\acro{MRC}{Maximum Ratio Combining}
\acro{MSR}{Multi-Stratum Resources}
\acro{MVNO}{Mobile Virtual Network Operator}
\acro{MZI}{Mach-Zehnder Interferometer}
\acro{MZM}{Mach-Zehnder Modulator}
\acro{NFV}{Network Function Virtualization}
\acro{NFVaaS}{Network Function Virtualization as a Service}
\acro{NFV-O}{Network Functions Virtualisation Orchestrator}
\acro{NG-PON2}{Next-Generation Passive Optical Network 2}
\acro{NLOS}{None Line Of Sight}
\acro{NMS}{Network Management System}
\acro{NRZ}{Non Return-to-Zero}
\acro{OBPF}{Optical BandPass Filter}
\acro{OBSAI}{Open Base Station Architecture Initiative}
\acro{OFC}{Optical Frequency Comb}
\acro{OFDM}{Orthogonal Frequency Division Multiplexing}
\acro{OLT}{Optical Line Termination}
\acro{ONAP}{Open Network Automation Platform}
\acro{ONF}{Open Networking Foundation}
\acro{ONU}{Optical Network Unit}
\acro{OOK}{On-off Keying}
\acro{OpenCord}{Central Office Re-Architected as a Data Center}
\acro{OPEX}{OPerating EXpense}
\acro{OSI}{Open Systems Interconnection}
\acro{OSS}{Operations Support Systems}
\acro{OXM}{OpenFlow Extensible Match}
\acro{PAM}{Pulse Amplitude Modulation}
\acro{PAPR}{Peak-to-Average Power Ratio}
\acro{PCE}{Path Computation Elements}
\acro{PCEP}{Path Computation Element Protocol}
\acro{PCF}{Photonic Crystal Fiber}
\acro{PD}{Photodiode}
\acro{PDCP}{RD Control Protocol}
\acro{PGW}{Packet Gateway}
\acro{PHY}{physical Layer}
\acro{PIR}{Peak Information Rate}
\acro{PMD}{Polarization Division Multiplexing}
\acro{PON}{Passive Optical Network}
\acro{PTP}{Precision Time Protocol}
\acro{PWM}{Pulse Width Modulation}
\acro{QAM}{Quadrature Amplitude Modulation}
\acro{QoE}{Quality of Experience}
\acro{QoS}{Quality of Service}
\acro{QPSK}{Quadrature Phase Shift Keying}
\acro{RAN}{Radio Access Network}
\acro{R-CORD}{Residential CORD}
\acro{RF}{Radio Frequency}
\acro{RLC}{Radio Link Control}
\acro{RN}{Remote Node}
\acro{ROADM}{Reconfigurable Optical Add Drop Multiplexer}
\acro{RoF}{Radio-over-Fiber}
\acro{RRH}{Remote Radio Head}
\acro{RRC}{Radio Resource Control}
\acro{RRPH}{Remote Radio and PHY Head}
\acro{RRU}{Remote Radio Unit}
\acro{RSOA}{Reflective Semiconductor Optical Amplifier}
\acro{Rx}{receiver}
\acro{SD-RAN}{Software Defined Radio Access Network}
\acro{SDMA}{Semi-Distributed Mobility Anchoring }
\acro{SDN}{Software Defined Network}
\acro{SDR}{Software Defined Radio}
\acro{SFBD}{Single Fiber Bi-Direction}
\acro{SGW}{Serving Gateway}
\acro{SIMO}{Single Input Multiple Output}
\acro{SMF}{Single Mode Fiber}
\acro{SNR}{Signal-to-Noise Ratio}
\acro{Split-PHY}{Split Physical Layer}
\acro{TED}{Traffic Engineering Database}
\acro{TEID}{Tunnel endpoint identifier}
\acro{TDD}{Time Division Duplex}
\acro{TD-LTE}{Time Division LTE}
\acro{TDM}{Time Division Multiplexing}
\acro{TDMA}{Time Division Multiple Access}
\acro{TWDM}{Time and Wavelength Division Multiplexing}
\acro{Tx}{transmitter}
\acro{UD-CRAN}{Ultra-Dense Cloud Radio Access Network}
\acro{UDP}{User Datagram Protocol}
\acro{UE}{User Equipment}
\acro{UFMC}{Universally Filtered Multicarrier}
\acro{UF-OFDM}{Universally Filtered OFDM}
\acro{UL}{Uplink}
\acro{USRP}{Universal Software Radio Peripheral}
\acro{vBBU}{virtualized BBU}
\acro{vBS}{virtual Base station}
\acro{vCPE}{virtual CPE}
\acro{VDSL2}{Very-high-bit-rate digital subscriber line 2}
\acro{VLAN}{Virtual Local Area Network}
\acro{VNF}{Virtual Network Functions}
\acro{VNTM}{Virtual Network Topology Manager}
\acro{VPE}{virtual Provider Edge}
\acro{VM}{Virtual Machine}
\acro{vOLT}{virtual OLT}
\acro{WAN}{Wide Area Network}
\acro{WAP}{Wireless Access Point}
\acro{WDM}{Wavelength Division Multiplexing}
\acro{WDM-PON}{Wavelength Division Multiplexing - Passive Optical Network}
\acro{WiMAX}{Worldwide Interoperability for Microwave Access}
\acro{WRPR}{Wired-to-RF Power Ratio}
\acro{XCI} {Crosshaul Control Infrastructure}
\acro{XFE} {Crosshaul Packet Forwarding Element}
\acro{XGPON}{10 Gigabit PON}
\acro{10GEPON}{10 Gigabit EPON}
\acro{XOS}{XaaS Operating System}
\acro{5GPoA}{5G Points of Attachments}

\end{acronym}


\end{document}